\documentclass[12pt]{article}
\usepackage{graphicx}
\newcommand\pubnumber{SNSN-323-63}
\newcommand\pubdate{\today}

\def\vtech{$^a$Center for Neutrino Physics, Department of Physics, Virginia Tech,\\
Blacksburg, VA 24061, USA}
\def\fukuoka{$^b$Department of Applied Physics, Fukuoka University,\\
Nanakuma 8-19-1, Fukuoka 814-0180, JAPAN}
\def\naoj{$^c$National Astronomical Observatory of Japan, \\
Mitaka, Tokyo 181-8588, JAPAN}
\def\support{\footnote{Poster presenter, email: nakamurako@fukuoka-u.ac.jp}}

\def\Title#1{\begin{center} {\Large #1 } \end{center}}
\def\Author#1{\begin{center}{ \sc #1} \end{center}}
\def\Address#1{\begin{center}{ \it #1} \end{center}}

\newcommand\pubblock{\rightline{\begin{tabular}{l} \pubnumber\\
         \pubdate  \end{tabular}}}
\newenvironment{Abstract}{\begin{quotation}  }{\end{quotation}}
\newenvironment{Presented}{\begin{quotation} \begin{center} 
             PRESENTED AT\end{center}\bigskip 
      \begin{center}\begin{large}}{\end{large}\end{center} \end{quotation}}
\def\Acknowledgements{\bigskip  \bigskip \begin{center} \begin{large}
             \bf ACKNOWLEDGEMENTS \end{large}\end{center}}
\def\beq{\begin{equation}}
\def\eeq#1{\label{#1}\end{equation}}
\def\eeqn{\end{equation}}

\def\beqa{\begin{eqnarray}}
\def\eeqa#1{\label{#1}\end{eqnarray}}
\def\eeqan{\end{eqnarray}}

\let\bar=\overbar

\def\Dslash{\not{\hbox{\kern-4pt $D$}}}
\def\dslash{\not{\hbox{\kern-2pt $\del$}}}

\def\msb{{\bar{\ssstyle M \kern -1pt S}}}

\begin{document}
\begin{titlepage}
\pubblock

\vfill
\Title{Diagnosing the Structure of Massive Stars\\with Galactic Supernova Neutrinos}
\vfill
\Author{Shunsaku Horiuchi$^a$, Ko Nakamura$^{b,}$\support, Tomoya Takiwaki$^c$, Kei Kotake$^b$}
\Address{\vtech \\ \fukuoka \\ \naoj}
\vfill
\begin{Abstract}
It has been suggested that whether a star explodes or not, and what kind of explosion properties it shows, is strongly dependent on the progenitor's core structure.
We present the results from 101 axisymmetric core-collapse supernova simulations performed with progenitors spanning initial masses in the range from 10.8 to 75 solar masses, and focus on their connections to the compactness of the progenitor's core. 
Our simulations confirm a correlation between the neutrinos emitted during the accretion phase and the progenitor's compactness. 
We suggest that the ratio of observed neutrino events during the first hundreds of milliseconds can be used to infer the progenitor's inner mass density structure. 
\end{Abstract}
\vfill
\begin{Presented}
NuPhys2017, Prospects in Neutrino Physics\\
Barbican Centre, London, UK,  December 20--22, 2017
\end{Presented}
\vfill
\end{titlepage}
\def\thefootnote{\fnsymbol{footnote}}
\setcounter{footnote}{0}

\section{Introduction}

Recently, simulations of core collapse based on large numbers of progenitor models have been performed. 
One-dimensional spherically symmetric studies have demonstrated that, in the neutrino-driven delayed explosion mechanism, explosion properties such as the explosion energy, synthesized nickel mass, and remnant mass, as well as neutrino luminosity and average energy, change non-monotonically with zero-age main sequence (ZAMS) mass of the progenitors~\cite{ugliano12,oconnor13}. 
Instead, they can be characterized by the compactness parameter which captures the density profile of the matter surrounding the collapsing core. 
These trends have been observed also in systematic two-dimensional axe-symmetric simulations~\cite{nakamura15}.

\begin{figure}[hb]
\centering
\includegraphics[height=3in]{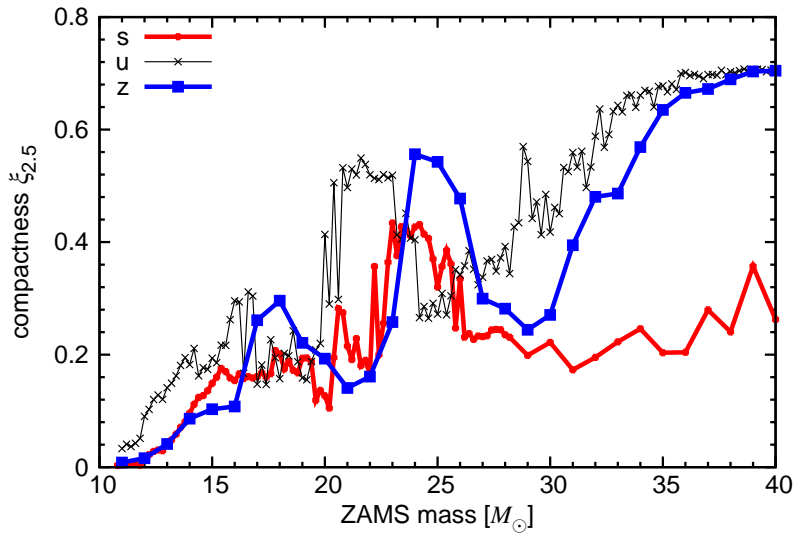}
\caption{Compactness parameter $\xi_{2.5}$ of core-collapse progenitor models from Woosley, Heger \& Weaver (2002) ~\cite{whw02}. 
Models with different metallicity (solar, ultra metal poor, and zero) are shown with different colors and symbols. 
It can be seen that the compactness is non-monotonic as a function of ZAMS mass.
}
\label{fig:xi}
\end{figure}

The compactness parameter is defined as
\begin{equation}
\xi_M = \frac{M/M_\odot}{R(M)/1000{\rm km}},
\end{equation}
where $R(M)$ is the radial coordinate that encloses mass $M$.
In this study we define $\xi_M$ at $M=2.5M_\odot$ ($\xi_{2.5}$) using pre-collapse progenitor profiles. %from Woosley, Heger \& Weaver (2002) ~\cite{WHW02}. 
Figure~\ref{fig:xi} presents the compactness parameter of progenitor models employed in the two-dimensional simulations by Nakamura et al. (2015)~\cite{nakamura15}.

%%%%%%%%%%%%%%%%%%%%%%%%%%%%%%%%%%%%%%%%%%%
%%%%%%%%%%%%%%%%%%%%%%%%%%%%%%%%%%%%%%%%%%%

\section{Core-Collapse Supernova Models}
We use the two-dimensional core-collapse models from Nakamura et al. (2015)~\cite{nakamura15}. 
Since our target is the future core-collapse supernova in our Galaxy, we focus on the solar-metallicity 101 models covering ZAMS mass from $10.8 M_\odot$ to $75 M_\odot$. 
These models are computed on a spherical polar grid of 384 radial zones from the center up to 5000 km and 128 angular zones covering $0 \leq \theta \leq \pi$. 
The equation of state by Lattimer \& Sweaty (1991)~\cite{lattimer91} for a compressibility modulus of $K=220$ MeV is employed with the energy feedback from nuclear reactions via 13 $\alpha$-nuclei network calculation.
The isotropic diffusion source approximation (IDSA)~\cite{idsa} with a ray-by-ray approach is used to solve the spectral transport of electron and anti-electron neutrinos. 
The cooling processes via heavy-lepton neutrinos are treated by means of a leakage scheme.

\begin{figure}[hb]
\centering
\begin{tabular}{cc}
\includegraphics[height=2.1in]{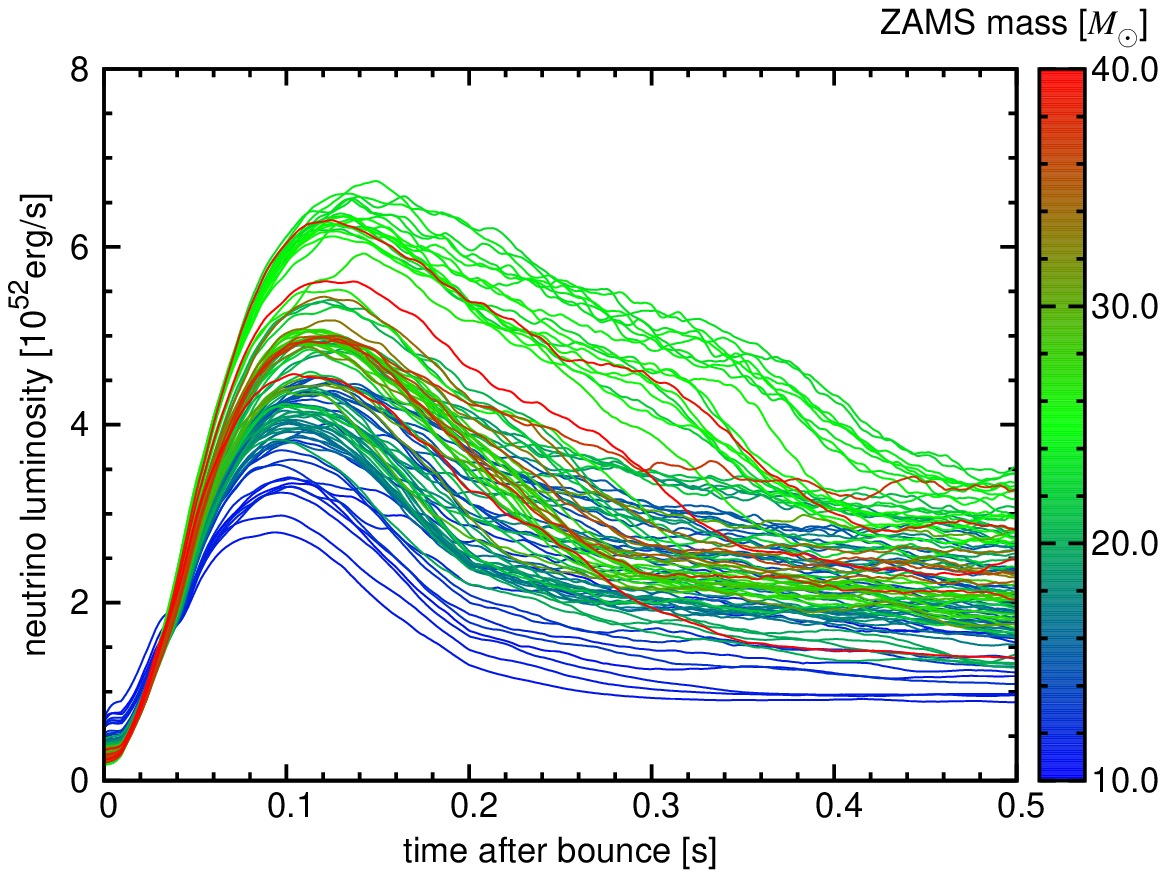} &
\includegraphics[height=2.03in]{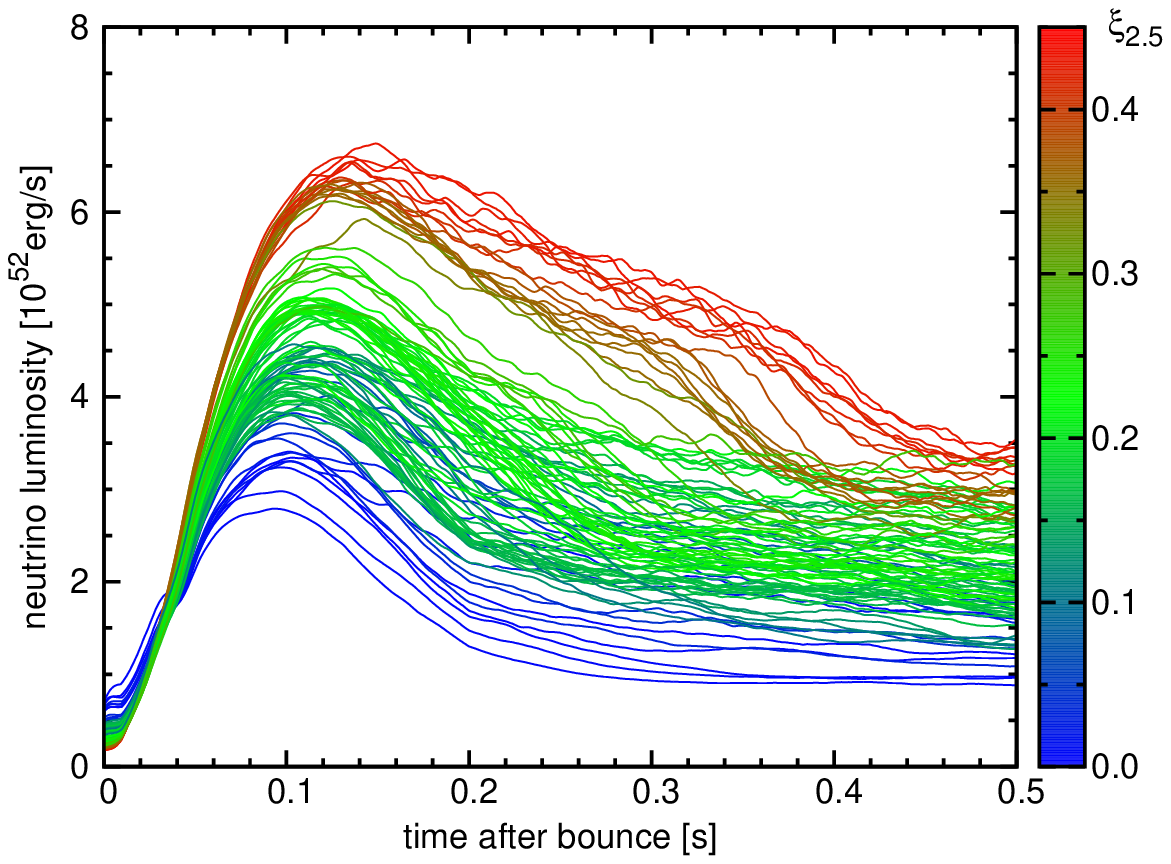}
\end{tabular}
\caption{Anti-electron neutrino luminosities as a function of time after bounce. Shown are 101 models with solar metallicity colored by their zero-age main sequence (ZAMS) mass ({\it left panel}) and by the compactness parameter $\xi_{2.5}$ ({\it right panel}).
The sudden drop of neutrino luminosities, for example at $\sim 0.35$ s after bounce for the models with the highest neutrino luminosity, is caused by shock revival and disappearance of accretion neutrino luminosity.}
\label{fig:lnueb}
\end{figure}

Figure~\ref{fig:lnueb} shows the time evolution of anti-electron neutrino luminosity for the examined 101 models. 
There is a wide variety among the models reflecting the different density profile of their progenitors. 
It is obvious that the compactness characterizes the model-dependent neutrino luminosity better than ZAMS mass since the neutrino luminosity in the early phase (a few milliseconds after bounce) is dominated by the neutrinos  from accreting matter and the so-called accretion luminosity is tightly correlated to a mass accretion rate onto the central core, which is well captured by the compactness parameter $\xi_M$ with a suitable choice of $M$.

%%%%%%%%%%%%%%%%%%%%%%%%%%%%%%%%%%%%%%%%%%%
%%%%%%%%%%%%%%%%%%%%%%%%%%%%%%%%%%%%%%%%%%%

\section{Estimating the compactness from neutrino detection}
The monotonic dependence of the neutrino luminosity on the compactness gives us an idea that the density structure of core-collapse supernova progenitor can be profiled by supernova neutrino. 
The problem is that the comparison of the neutrino detection events and supernova neutrino models involves systematic uncertainties such as the distance to the supernova. 
To overcome this problem, we propose a new indicator which is independent of the distance but sensitive to the compactness. 
Figure~\ref{fig:lnueb} tells us that the neutrino luminosity during the first 50 msec is nearly independent of the compactness, while later time windows, for example 200--250 msec, show a strong dependence of the compactness.
Therefore, the ratio of the detection events between two time windows, $N_{200-250{\rm ms}}/N_{0-50{\rm ms}}$, 
keeps the compactness dependence and cancels the distance uncertainty. 

\begin{figure}[htb]
\centering
\includegraphics[height=4in]{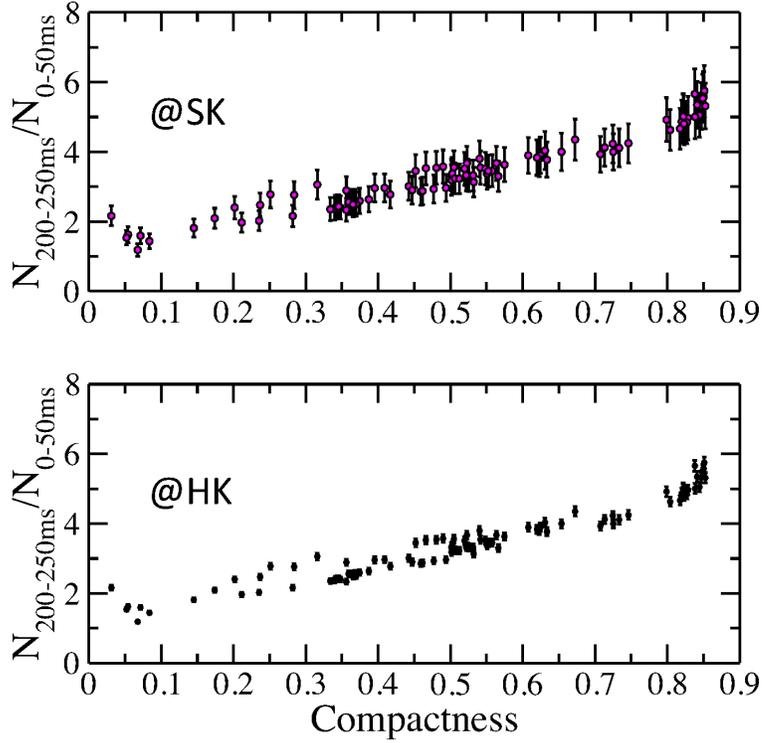}
\caption{Predicted ratio of anti-electron neutrino detection events at Super-Kamiokande ({\it top panel}) and Hyper-Kamiokande ({\it bottom panel}) as a function of compactness $\xi_{2.5}$. 
All points include statistical errors. 
Here we adopt MSW mixing under normal mass hierarchy and the distance of 10 kpc.}
\label{fig:nuevent}
\end{figure}

Figure~\ref{fig:nuevent} shows the result. 
We find an monotonically increasing trend of the ratio as a function of the compactness. 
A large neutrino detector such as Hyper-Kamiokande is capable of estimating the progenitor compactness of a Galactic supernova.

%%%%%%%%%%%%%%%%%%%%%%%%%%%%%%%%%%%%%%%%%%%
%%%%%%%%%%%%%%%%%%%%%%%%%%%%%%%%%%%%%%%%%%%

\section{Summary and Discussions}
We have presented a simple way of using neutrinos to probe the core structure of supernova progenitor stars. 
A simple ratio of neutrino detection event rates will be useful to reveal the collapsing core has a large compactness or not. 
There remain some uncertainties to be discussed, for example, the dimensionality of supernova models, supernova neutrino treatment, and the impact of additional neutrino flavor mixing beyond MSW. 
More detailed analysis with more sophisticated supernova neutrino models is demonstrated in our latest paper~\cite{horiuchi17}, although the number of models is small.

\Acknowledgements
This study was partly supported by JSPS KAKENHI Grant Numbers 26104007, JP16K17668, and JP17H05205. 
Numerical computations were in part carried out on Cray XC30 at Center for Computational Astrophysics, National Astronomical Observatory of Japan.
KN is also supported by funds from the Central Research Institute of Fukuoka University (Nos. 171042, 177103).

\end{document}